\documentclass[11pt]{article}
\usepackage{amsmath}
\usepackage{latexsym}
\usepackage{amssymb}
\usepackage{fullpage}
\usepackage{tikz}
 \usepackage{colortbl}
\usepackage{epsfig}
\usepackage{graphics}

\def\dimo{\noindent\mbox{\sc proof: }}
\def\eproof{\rm\hspace*{\fill}$\Box$\vspace{10pt}}
\newtheorem{defin}{\bf Definition}[section]
\newtheorem{theo}[defin]{Theorem}

\newtheorem{prop}[defin]{Proposition}

\newtheorem{oss}[defin]{Remark}
\newtheorem{ex}[defin]{Example}

\def\cW{{\cal W}}

\def\cP{\mbox{\boldmath ${\cal P}$}}
\def\cS{{\cal S}}

\def\cO{{\cal O}}

\def\meglio {{ {\underset {\sim}  \succ}}}

\def\cP{\mbox{\boldmath ${\cal P}$}}
\def\cS{{\cal S}}

\title{{\sc {\Large On the relation between Preference Reversal and Strategy-Proofness
}}}
\author{\normalsize K. P. S. Bhaskara Rao\\
{\small Department of Computer Information Systems}\\
{\small Indiana 
University Northwest, Gary, IN 46408}\\
{\small E-mail: bkoppart@iun.edu}\\
\normalsize Achille Basile\thanks{Corresponding author.}\\
{\small Dipartimento di Scienze Economiche e Statistiche} \\
{\small Universit\`a Federico II,
80126 Napoli, Italy}\\
{\small E-mail: basile@unina.it}
\\
\normalsize Surekha Rao \\
{\small School of Business and Economics}\\
{\small Indiana University Northwest,
Gary, IN 46408}\\
{\small E-mail: skrao@iun.edu}
}

\begin{document}
\maketitle

\thispagestyle{empty}

\begin{abstract}
 We analyze the relation between strategy-proofness and preference reversal  in the case that agents may declare indifference. Interestingly, Berga and Moreno (2020), have recently derived preference reversal from group strategy-proofness of social choice functions on strict preferences domains if the range has no more than three elements. 
We extend this result and at the same time simplify it. Our analysis points out the role of individual strategy-proofness in deriving the preference reversal property, giving back to the latter its original individual nature (cfr. Eliaz, 2004).
   
     Moreover, we show that the difficulties Berga and Moreno highlighted relaxing the assumption on the cardinality of the range, disappear under a proper assumption on the domain.  We introduce  the concept of  complete sets of preferences and show that  individual strategy-proofness is sufficient to obtain the preference reversal property when the agents' feasible set of orderings is complete. This covers interesting cases like single peaked preferences, rich domains admitting regular social choice functions, and universal domains. The fact that we  use individual rather than group strategy-proofness,    allows to get immediately some of the known, and some new, equivalences between individual and group strategy-proofness. Finally, we show that group strategy-proofness is only really needed to obtain preference reversal if there are infinitely many voters. 
\end{abstract}

JEL Code: D71

Mathematics Subject Classification: 91B14

{\it{Keywords:
Social choice function, Preference reversal,
Strategy-proofness. }}

 \section{Introduction}
In a framework where individuals from a society $V=\{1, 2, \dots, n\}$ express strict preferences on a set $A=\{a, b, c, \dots\}$ of alternatives, Berga and Moreno \cite{BM} recently conducted  a study on the relation between group strategy-proofness and a property  of a social choice function, namely  preference reversal, introduced by Eliaz \cite{E}. 

Group strategy-proofness is a well known strengthening of the usual (i.e. individual) strategy-proofness: it  prevents a  social choice rule from being manipulated  by groups of individuals
 (see Barber\`a et al. \cite{BBM} for a sharp distinction between group and individual strategy-proofness). Preference reversal has been introduced, for strict preference orders, as a property  of a social choice function that looks rather natural and   innocuous, one is tempted to say. A social choice function satisfies the preference reversal property if: given that $\,a\,$ is the social choice corresponding to a profile $P$ of preferences and $\,b\,$ is  chosen instead if a different profile $Q$ prevails, then at least one individual that was preferring $a$ to $b$ in the profile $P$ must have moved to prefer $b$ to $a$ under $Q$.

The interesting finding of Berga and Moreno is that   group strategy-proofness is implied by the preference reversal property \cite[Theorem 3]{BM} and that the two properties are equivalent in case the set $A$ has at most three alternatives \cite[Theorem 4]{BM}. The equivalence does not hold true in general, as they show in \cite[Example 5]{BM}

Our purpose in this short note is twofold. First, in the case 
 that agents may declare indifference, a model more general than the one considered by Berga and Moreno \cite{BM}, we formulate a definition of the property of preference reversal and, through a simple and short induction  argument,  we show the direct role played  by the individual strategy-proofness in obtaining  the conclusion of \cite [Theorem 4]{BM}. To be precise we show that: {\it if the range of a social choice function has cardinality at most three, then individual strategy-proofness is sufficient to obtain the preference reversal property.} The Berga and Moreno \cite [Theorem 4]{BM} is a consequence of our result. The Barber\`a et al. \cite[Corollary 1]{BBM}, according to which  individual and group  strategy-proofness  are equivalent when the range of a social choice function has cardinality at most three, also follows from  result.
  In our framework, we also introduce the notion of  almost preference reversal property, and observe that it is equivalent to group strategy-proofness.

 \medskip

Secondly, we show that the difficulties highlighted by \cite[Example 5]{BM} when the range of the social choice function has cardinality more than 3, disappear under an appropriate assumption on the domain. We introduce  the concept of  complete sets of preferences and show that {\it individual strategy-proofness is sufficient to obtain the preference reversal property when the agents' feasible set of preferences is complete.} This covers the cases of universal domain, singe peaked preferences, and the case of rich domains admitting regular social choice functions (\cite{LZ} and \cite{BBM}).
Finally, we describe a simple trick and show that our results can be extended to infinite sets of individuals when there are finitely many alternatives in $A,$ and group strategy-proofness is assumed. This marks a distinction between group and individual strategy-proofness as a sufficient condition for the preference reversal property.

\bigskip
The paper proceeds  by presenting the model in Section 2,  the results  and their discussion in Section 3.  Conclusions are  presented in Section \ref{conclusioni}.

\section{The Model}

Let $A$ and $V$ be arbitrary sets that represent, respectively, the set of the available alternatives $\,\, a,b,c,\dots \,\,$ and the set of agents $\, v\,$. Let $\cW$ be the set of weak orderings, or weak preferences, or simply preferences, for short, over the set $A$. Its subset consisting of strict orderings over $A$ will be denoted by $\cS$. 
 With reference to   an element  $W\in \cW$, the notation $ x\quad {\underset {\sim}  \succ}_{W} y$  stands for $(x,y)\in W$, the notation $ x\, \underset W{\succ} \,y$ stands for $\big[(x,y)\in W$ and $(y,x)\notin W\big]$, the notation $ x\, \underset W{\sim} \,y$ stands for $\big[(x,y)\in W$ and $(y,x)\in W\big]$.

A preference profile is  $P=(P_v)_{v\in V}$ where $P_v\in \cW$ for every voter $v \in V$.

Let $\cP$ be a subset of the class $\cW^V$ of all preference profiles. A social choice function (scf, for short) is a mapping
\,\,
$\phi: \cP\to A.$\,
The model assumes 
that for every agent $v$ there is a feasible set $\cW_v\subseteq \cW$ of preference relations to be used in contributing to the formation of a preference profile. In other words we assume that $\cP$ is a cartesian product

$$\cP=\{ P\in \cW^V: P_v\in \cW_v, \, \forall v\in V\}=\underset {v\in V}\times \, \cW_v,$$
the sets $\cW_v$ of preferences being primitives of the model.

The literature refers to the special cases 

- $\cW_v=\cW$ for all $v\in V$,  and 

- $\cW_v=\cS$ for all $v\in V$ 

as to the Universal Domain Hypothesis (one may wish to add the distinction weak or strict, with obvious meaning). The, indeed more general, case we consider here is referred to as the Restricted Domain Hypothesis.

For a profile $P=(P_v)_{v\in V}$ we shall also use the notation  $P=[P_I, P_{I^c}]$   if $I$ is a subset of $V$ and $P_I$, $P_{I^c}$ are the obvious restrictions $P_I=(P_v)_{v\in I}$, $P_{I^c}=(P_v)_{v\notin V}$ of $P$\footnote{We denote by the superscript $^c$ the complement of a set.}. 

We also shorten $\phi([P_I, P_{I^c}])$ as $\phi(P_I, P_{I^c})$, and, if the set $I$ consists of one player $\{v\}$,  we write $P_v$ and $P_{-v}$ instead of $P_{\{v\}}$ and $P_{\{v\}^c}$, as it is usual.

\def\alf{either $\left\{ \begin{array} {ll}
\phi(P)&\underset{P_v}\succ \phi(Q)  \\
\phi(Q)&\meglio_{Q_v} \phi(P)
\end{array}
\right.$ is true, or 
$\left\{ \begin{array} {ll}
\phi(P)&\meglio_{P_v} \phi(Q)  \\
\phi(Q)&\underset{Q_v}\succ \phi(P)
\end{array}
\right.$ is true}

\def\bet{$\left\{ \begin{array} {ll}
\phi(P)&\meglio_{P_v} \phi(Q) \,\,\,\meglio_{Q_v} \phi(P) \\
P_v&\neq \,Q_v
\end{array}
\right.$}

\def\bet{$\left\{ \begin{array} {ll}
\phi(P)&\,\meglio_{P_v} \phi(Q) \\
\phi(Q)&\,\meglio_{Q_v} \phi(P) \\
P_v&\neq \,Q_v
\end{array}
\right.$}
 
 \def\gam{$\left\{ \begin{array} {ll}
\phi(P)&\meglio_{P_v} \phi(Q)  \\
P_v&\neq \,Q_v
\end{array}
\right.$}

\def\alfab{either $\left\{ \begin{array} {ll}
a&\underset{P_v}\succ b  \\
b&\meglio_{Q_v} a
\end{array}
\right.$ is true, or 
$\left\{ \begin{array} {ll}
a&\meglio_{P_v} b  \\
b&\underset{Q_v}\succ a
\end{array}
\right.$ is true}

\def\betab{$\left\{ \begin{array} {ll}
a&\meglio_{P_v} b  \\
b&\meglio_{Q_v} a \\
P_v&\neq \,Q_v
\end{array}
\right.$}

\def\gamab{$\left\{ \begin{array} {ll}
a&\meglio_{P_v} b  \\
P_v&\neq \,Q_v
\end{array}
\right.$}

\bigskip

We now introduce the notion of Preference Reversal property in the case of weak orders. 
\begin{defin}\label{PR3}

A scf $\phi$ satisfies the {\bf Preference Reversal} property if:

$\phi(P)\neq\phi(Q) \implies \exists v\in V$ such that \bet

\end{defin}

When individuals are not allowed to declare indifference, i.e. $\cW_v\subseteq \cS$ for all $v\in V$, then  Definition \ref{PR3} coincides with \cite[Definition 2]{BM}.

The notion of almost preference reversal property is introduced next with the aim of  characterizing  group strategy-proofness, as the  Proposition \ref{APR=GSP} shows.

\begin{defin}\label{APR}
A scf $\phi$ satisfies the {\bf Almost Preference Reversal} property if:

$\phi(P)\neq\phi(Q) \implies \exists v\in V$ such that \gam
\end{defin}

Note that Almost Preference Reversal Property requires only that there exist two voters $v$ and $w$ such that $\phi(P)\,\, \meglio_{P_v} \phi(Q)$  and $\phi(Q) \,\,\meglio_{Q_w} \phi(P)$, and that $v$ need not be same as $w.$ In the Preference Reversal property we are demanding that there is a unique $v$ that satisfies both the properties.

\bigskip
Let $\phi$ be scf, $D$ a coalition of voters, and $P\in \cP$ a profile of preferences.
 We recall  that   if there is a partial  profile $Q_D=(Q_v)_{v\in D}$, with $Q_v\in \cW_v$, such that 
every individual $v$ of the coalition D  prefers $\phi(Q_D, P_{D^c})$ to $\phi(P)$ according to $P_v$, \, i.e. $\phi(Q_D, P_{D^c})\underset{P_v}\succ \, \phi(P),\,  \forall v\in D$, then  one says that the coalition $D$ can manipulate   $P$   under $\phi$
 by presenting the profile $Q_D$.

\begin{defin}\label{GSP}
 A scf $\phi$ is said to be  {\bf  {\color{black}group} strategy-proof} (GSP, for short) if no coalition of voters can manipulate any profile under $\phi$. 
\end{defin}

If in the definition of GSP we replace arbitrary coalitions with singleton sets, we have the notion  of
{\bf individual strategy-proofness} ({\it ISP}, for short).

\section{Results}

We start by showing that almost preference reversal property characterizes  group strategy-proofness.

Evidently, the definition of group strategy-proofness can be reformulated by requiring the validity of the following implication for a scf $\phi$:
$$
\big[ D \mbox{ is a coalition, } P, Q\in \cP \mbox{ are profiles identical on } D^c   \big] \Rightarrow \big[ \exists v\in D: \,\phi(P) \meglio_{P_v} \phi(Q)    \big].
$$

 and this allows us to state the following proposition.

\begin{prop}\label{APR=GSP} 
A scf is group strategy-proof if and only if it satisfies the almost preference reversal property.
\end{prop} 
\dimo
That group strategy-proofness gives the almost preference reversal property, as a
matter of fact, has been already observed above.
 
For the converse, let us negate  GSP for  $\phi$: for a coalition $D$, two profiles $P$ and $Q=[Q_D, P_{D^c}]$ we have $\phi(Q_D, P_{D^c})\underset{P_v}\succ \phi(P), \, \forall v\in D$. If $\phi$ satisfies the almost preference reversal property, there is a voter $v$, necessarily in $D$ since the condition $P_v\neq Q_v$ is also assumed, with $\phi(P)\meglio_{P_v}\phi(Q)$, a contradiction.
\eproof

 In \cite[Theorem 3]{BM} it was shown that group strategy-proofness follows from preference reversal property.  We have provided above the missing link in the sense that our definition of almost preference reversal property does in fact characterize group strategy-proofness.  Proposition \ref{APR=GSP}, when restricted to strict profiles, supplements  \cite[Theorem 3]{BM} with a converse result.
 
\bigskip

Evidently, the property of preference reversal gives the individual strategy-proofness of a scf. The following theorem provides the first of our results about reversing the implication. It generalizes  Berga and Moreno  \cite[Theorem 4]{BM} and, at the same time, provides a very short induction argument to show the validity of the result.
 \begin{theo}\label{main}
Let $\phi$ be a scf. If $V$ is finite and the range  of $\phi$ has cardinality at most 3, then:

$\phi$  is ISP $\Rightarrow \phi$ satisfies the Preference Reversal property.  
\end{theo}
\dimo
We prove the theorem by induction on the cardinality of the society $V$.  Clearly the theorem is true when $|V| = 1.$ So let us suppose that the theorem is true when $|V|=n$ and prove that it is also true when $|V|=n+1$. 

Assume $a=\phi(P)\neq\phi(Q)=b$ with $P,Q\in \cW_1\times\dots\times\cW_n\times\cW_{n+1}$.
Take  a voter  $v$ for which $P_v\neq Q_v.$

Since the range of our scf is at most of cardinality three, $\phi(P_v, Q_{-v})$  can be either $a,$ or $b$ or, possibly, a third value $c.$

According to the values of $\phi(P_v, Q_{-v})$ we discuss the following cases:

Case 1:  $\phi(P_v, Q_{-v})=a.$ 
We also know that $\phi(Q_v, Q_{-v})=b.$ An application of ISP property of $\phi$ gives us that $a \,\,\meglio_{P_v} b$  and $b\,\, \meglio_{Q_v} a$.

Case 2:  $\phi(P_v, Q_{-v})=b.$
We also know that $\phi(P_v, P_{-v})=a.$ We consider the function $\phi(P_v, \ldots).$ This is ISP. Hence by the induction hypothesis there is voter $w \in V \setminus \{v\}$ that witnesses preference reversal.

Case 3:  $\phi(P_v, Q_{-v})=c.$ We will divide this into three sub-cases according to the values of $\phi(Q_v, P_{-v}).$ If we have 
  $\phi(Q_v, P_{-v})=a,$ we consider the function $\phi(Q_v, \ldots)$ and use induction as in Case 2 above.

The subcase  ( $\phi(P_v, Q_{-v})=c$ and ) $\phi(Q_v, P_{-v})=b,$  is similar to Case 1  above since $\phi(P_v, P_{-v})=a.$

Finally we have to analyze the subcase:  $\phi(P_v, Q_{-v})=c$ and  $\phi(Q_v, P_{-v})=c.$

Since $\phi(P_v, Q_{-v})=c$ and $\phi(Q_v, Q_{-v})=b,$ by ISP we have $c\,\,\meglio_{P_v} b$  and $b\,\, \meglio_{Q_v} c$. 

Since $\phi(Q_v, P_{-v})=c$ and $\phi(P_v, P_{-v})=a,$ by ISP we have $c\,\,\meglio_{Q_v} a$  and $a\,\, \meglio_{P_v} c$. 
By transitivity, we see that $a\,\,\meglio_{P_v} b$  and $b\,\, \meglio_{Q_v} a$.  Thus we have preference reversal.
\eproof

Comparing with \cite[Theorem 4]{BM}, we see that our  result 

- clearly points out the  individual nature of the Preference Reversal property,

- it is given for the more general setting of weak orderings,

-  and at the same time presents a  very simple  proof.

 If we consider only strict orderings as in \cite{BM}, formally \cite[Theorem 4]{BM} and Theorem \ref{main} can be seen as equivalent. The latter, indeed, can be obtained from the former because when the range of $\phi$ has cardinality at most three, it is known from \cite[Corollary 1]{BBM} that $\phi\,\,   ISP \Rightarrow \phi \,\, GSP$.  Yet, as discussed in Remark \ref{final}, our direct approach gives some insights about the implication $\phi\,\, ISP \Rightarrow \phi \,\, GSP$, and in particular \cite[Corollary 1]{BBM} is a corollary of Theorem \ref{main}.

\bigskip
\begin{oss}\label{foot GS}
 {\rm Theorem \ref{main} cannot be extended without the restriction on the range of the scf. This is clarified by the example  \cite[Example 5]{BM} of Berga and Moreno. However, if we assume some properties for the domain of the scf,  we can give up on the condition on the range of $\phi.$ 
 
 The simplest possibility is to assume the Universal Domain hypothesis.
 Certainly,  the Preference Reversal property is a trivial consequence of the Gibbard-Satterthwaite Theorem if one remains confined to use strict preferences. However dealing with weak orderings one must be a bit more accurate in the verification that an ISP scf satisfies the Preference Reversal property.{\footnote{The following would be a possible argument. If we suppose that: $\phi: \cW^{n}\to A$ is ISP, its range has more than three elements,  and $a=\phi(P)\neq\phi(Q)=b$, let $v$ be a dictator. By definition {\color{black} (\cite[Definition 6.4]{JR})} :
$a$ is one of the top elements for $P_v$ and $b$ is one of the top elements for $Q_v$. The conclusion that the Preference Reversal is satisfied would be achieved if  it were true that one has $P_v\neq Q_v$. But this cannot be guaranteed in general (unless the domain of $\phi$ is $\cS^n$ instead of $\cW^n$). So, let us suppose that this is not the case and $P_v=Q_v$. 
In such a case we could   repeat the same induction argument used in the proof of Theorem \ref{main} to obtain the Preference Reversal property. } } 

Remarkably, finer results can be obtained  in a surprisingly simple and  direct way, that is without appealing to the Gibbard-Satterthwaite Theorem. Again induction will be  used to achieve this.}
\end{oss}

Let us introduce some useful definitions, and provide illustrative examples.

\begin{defin}  Given a weak ordering $P$ and an alternative $a,$ we say that a weak ordering $W$ 
resolves  the $P$-indifference at $a$
 if 
$$a\,\,\meglio_{P} x, \, x\neq a\Rightarrow a\underset{W}\succ x.$$
\end{defin}

\begin{defin}
Given two alternatives 
$a\neq b,\in A$  and the weak orderings  $P, Q, W$, we say that $W$  is an  $(a,b)$-resolvent of $(P, Q)$ if $W$ at the same time resolves the $P$-indifference at $a$ and the $Q$-indifference at $b$.
\end{defin}
 Clearly  an $(a,b)$-resolvent of a pair of profiles $(P, Q)$ can only exist if the conditions $a\,\,\meglio_{P} b$ and $b\,\,\meglio_{Q} a$ are not simultaneously true.

\begin{ex}\label{35}
 {\rm 
Let  us denote by $W_{(a,b)}$  a weak ordering such that $a\underset{W}\succ b$ and all the other alternatives are, according to $W$, strictly below $b$, i.e. $W_{(a,b)}$ denotes an arbitrary element of the set $\cO_{(a,b)}=\{W\in\cW: a\underset W\succ b\underset W\succ A\setminus\{a, b\} \}.$

Let us suppose that the conditions \,\, $a\,\,\meglio_{P} b$ and $b\,\,\meglio_{Q} a$ \,\, are not simultaneously true.  Then we immediately see that:
 
 - if  $a\underset{Q}\succ b$, then
$W_{(a,b)}$   is a resolvent of $(P,Q)$, for all $P$.

- if  $b\underset{P}\succ a$, then
$W_{(b,a)}$   is a resolvent of $(P,Q)$, for all $Q$.
}\eproof
\end{ex}

\begin{ex}\label{35bis}
 {\rm 
 Let us suppose that the conditions \,\, $a\,\,\meglio_{P} b$ and $b\,\,\meglio_{Q} a$ \,\, are not simultaneously true, for given pairs $(a,b)$ of distinct alternatives, and $(P,Q)$ of preferences. If  $a\underset{Q}\succ b$, 
let  us denote by $W'(a, b, P, Q)$  a weak ordering such that, omitting $(a,b, P,Q)$ for simplicity below,
\begin{itemize}
\item the alternatives in the set $L(Q, b):=\{x\in A: x\neq b, b\,\,\meglio_Q x\}$ are $W'$-indifferent 

\item the alternatives in the set $L(P,a)\setminus L(Q, b)$ are $W'$-indifferent

\item $ a  \underset{W'}\succ L(P,a)\setminus L(Q, b) \underset{W'}\succ b  \underset{W'}\succ  L(Q, b)$

\end{itemize}
\bigskip
  Then we immediately see that
$W'(a, b, P, Q)$   is an $(a,b)$-resolvent of $(P,Q)$.

If  $b\underset{P}\succ a$, then $W'(b, a, Q, P)$   is an $(a,b)$-resolvent of $(P,Q)$.
}\eproof
\end{ex}

\begin{ex}\label{37}
 {\rm 
 Examples can be also presented with reference to single peaked preferences. 
Let us suppose that the set $A$ of alternatives is identified with the ordered set of the first $k$ natural numbers $1, 2, 3, \dots, k$.  Let $P, Q$ be  single peaked preferences and $a$ and $b$ be two alternatives. We will see that there is a single peaked $W$ that is an $(a,b)$-resolvent of $(P, Q)$.
Under the assumption that the conditions $a\,\,\meglio_{P} b$ and $b\,\,\meglio_{Q} a$ are not simultaneously true, a single peaked resolvent can be defined as follows.

\bigskip
In case $b\underset P\succ a$ a desired resolvent $W$ has $b$ as unique $W$-top element, and further, in case $b<a$,
\begin{itemize}

\item $a\underset  W\succ b -1$

\item $b\underset  W\succ b +1 \underset  W\succ b +2\underset  W\succ \dots \underset  W\succ a\underset  W\succ a+1\underset  W\succ \dots\underset  W\succ k$

\item $b\underset  W\succ b -1 \underset  W\succ b -2\underset  W\succ \dots \underset  W\succ 1$
\end{itemize}

Remaining in the case  $b\underset P\succ a$, but with $b>a$, $W$ has $b$ as unique $W$-top element, and further,
\begin{itemize}

\item $a\underset  W\succ b +1$

\item $b\underset  W\succ b +1 \underset  W\succ b +2\underset  W\succ \dots \underset  W\succ k$

\item $b\underset  W\succ b -1 \underset  W\succ b -2\underset  W\succ \dots \underset  W\succ a\underset  W\succ a-1\underset  W\succ a-2\underset  W\succ \dots\underset  W\succ 1$.
\end{itemize}
In case we have $a\underset Q\succ b$, the example we can provide is as above, just replacing the role of $a$ and $b$.\eproof
}\end{ex}

\begin{defin}\label{ED}
A  set $\cO\subseteq \cW$ of weak orderings is said to be {\bf complete} if it contains an $(a,b)$-resolvent $W$ of $(P, Q)$ whenever $P, Q\in\cO,\,a\neq b\in A, \, $ and the conditions $a\,\,\meglio_{P} b$ and $b\,\,\meglio_{Q} a$ are not simultaneously true.
\end{defin}

On the basis of Examples \ref{35}, \ref{35bis} and \ref{37},  and by direct comparison  with the definitions of 

- rich domains (\cite[Definition 6] {LZ}, \cite[Definition 9]{BBM}) 

- regular scfs (\cite[Definition 3] {LZ}, \cite[Definition 10]{BBM}) 

we can state what follows.

\begin{prop}
The following are complete sets of preferences:
\begin{itemize}
\item the set $\cW$, of all preferences (and so is the set $\cS$);
\item the set of all single-peaked preferences;
\item every set ${\cal O}$ of preferences such that for every $a \neq b$ a preference  $W_{(a, b)} \in \cal O$.
\item every set ${\cal O}$ of preferences such that for every pair $(a,b)$ of distinct alternatives and every pair $(P,Q)$ of preferences, then a preference  $W'(a, b, P, Q) \in \cal O$.
\item every rich set ${\cal O}$ of preferences such that $\cO^V$ is the domain of an onto, regular scf.
\end{itemize}
\end{prop}

When agents may select their preferences  from a complete subset of $\cW$, the restriction of no more than three alternatives in the range can be relaxed in order to obtain the preference reversal property from the strategy-proofness. This is the content of the next theorem.

\begin{theo}\label{New}
 Suppose $V$ is finite. Let $\phi: \cW_1\times\dots\times\cW_n\to A$ be an ISP scf.  If for every agent $v$ the feasible set $\cW_v$ of preferences is complete, then $\phi$ satisfies the Preference Reversal property.
\end{theo}
\dimo
We prove the theorem by induction on the cardinality of the society $V$.  Clearly the theorem is true when $|V| = 1.$ So let us suppose that the theorem is true when $|V|=n$ and prove that it is also true when $|V|=n+1$. 

Assume $a=\phi(P)\neq\phi(Q)=b$ with $P,Q\in \cW_1\times\dots\times\cW_n\times\cW_{n+1}$.

If for the first voter we have $P_1=Q_1$,  since the function $\phi(P_1, \dots)$ is ISP, by the induction hypothesis we shall get immediately a voter $v\ge 2$ that witnesses preference reversal. So we can suppose that $P_1\neq Q_1$. 
If the conditions $a\,\,\meglio_{P_1} b$ and $b\,\,\meglio_{Q_1} a$ are simultaneously true, then voter 1 itself witnesses preference reversal. If this is not the case,
let $W\in\cW_1$ be an $(a,b)$-resolvent of $(P_1, Q_1)$.

Since $\phi$ is ISP, then also $\phi(W, P_{-1})= a$. Let us show this. 

Suppose $\phi(W, P_{-1})= x$. 

For the value $x$ it is not possible that $x\underset{P_1}\succ a$, because we also have 
$\phi(W, P_{-1})= x$,  $\phi(P_1, P_{-1})= a$.
Therefore we necessarily have  $a\,\,\meglio_{P_1} x$, and if $x\neq a$, the resolvent properties say that $a\underset{W}\succ x$. This together with $\phi(W, P_{-1})= x$,  $\phi(P_1, P_{-1})= a$, violates ISP, hence the proof that $\phi(W, P_{-1})= a$ is concluded.

\bigskip

Also we can say that $\phi(W, Q_{-1})= b$. Indeed assume $\phi(W, Q_{-1})= x$.

\bigskip

For the value $x$ it is not possible that $x\underset{Q_1}\succ b$, because we also have 
$\phi(W, Q_{-1})= x$,  $\phi(Q_1, Q_{-1})= b$.

Therefore we necessarily have  $b\,\,\meglio_{Q_1} x$, and if $x\neq b$, the resolvent properties say that $b\underset{W}\succ x$. This together with $\phi(W, Q_{-1})= x$,  $\phi(Q_1, Q_{-1})= b$, violates ISP. 
So it remains only the possibility that $\phi(W, Q_{-1})= b$. Consequently, applying the induction hypothesis to $\phi(W, \dots)$ again we shall get immediately a voter $v\ge 2$ that witnesses preference reversal.
\eproof

\bigskip
After Proposition \ref{APR=GSP} and Theorems \ref{main} and \ref{New} we can summarize as follows our results up to now.
\begin{theo}
For a social choice function $\phi: \underset {v\in V}\times \, \cW_v\to A $,   the following is true:

\begin{itemize}
\item[(i)] $\phi$ satisfies the Preference Reversal property $\Rightarrow \phi$ satisfies the Almost Preference Reversal property $\Leftrightarrow \phi$ is GSP $\Rightarrow \phi$ is ISP.

\item[(ii)]
 If $V$ is finite and $\phi$ either takes at most three values or if  the feasible set of orderings of every voter  is complete,  then we  have further that:

 $\phi$ is ISP $\Rightarrow$ $\phi$ satisfies the Preference Reversal property,
 that is in this case  Preference Reversal, Almost Preference Reversal, Group Strategy-Proofness, Individual Strategy-proofness are equivalent properties of a scf. 
 \end{itemize}
\end{theo}

Since in our model agents may declare indifference, we cannot add, as a further equivalence, that $\phi$ is a dictatorial function (\cite[Definition 6.4]{JR}) as the following example shows.

\begin{ex}
{\rm Consider a society with voters $\{1, 2\}$ and alternatives $A=\{a, b,c\}$.
Define on the universal domain a scf $\phi$ as follows.

\medskip
{\it $\phi(P)$ is the $P_1$-top alternative, if such an alternative is unique. Otherwise among the $P_1$-top alternatives we select as 
$\phi(P)$,  the alphabetical first element of   the ${P_2}^{-1}$- top elements.}

\medskip
This scf is dictatorial by definition.  But it is not ISP \footnote{Indeed consider $P=[P_1, P_2], Q=[P_1, Q_2]$ where  $a\underset{P_1}\sim b\underset{P_1}\succ c$, $b\underset{P_2}\succ \, a$, $\,Q_2=P_2^{-1}$.}. Also, $\phi$ does not satisfy the Preference Reversal property. So the obviousness that a dictatorial scf satisfies the Preference Reversal property for the case of strict orderings, is not true when dealing with weak orderings. }
\end{ex}

The finiteness of $V$ plays a crucial role in the arguments given before. However, as we are going to show, the results can be extended  to infinite societies if the individual strategy-proofness is strengthened by assuming group strategy-proofness.  We will reduce, in the way shown by the proof of the following Theorem, the infinite case to the finite one.

\begin{theo}\label{main infinito}
Suppose $V$ is infinite and the set of alternatives is finite. Let $\phi$ be a GSP scf. Then, $\phi$ satisfies the Preference Reversal property if either one of the following hypothesis is assumed:

(1)   the range of $\phi$  is of cardinality at most three,

(2) every voter has the same complete  set $\cO$ of feasible orderings.

\end{theo}
\dimo 
Since we want to show that the Preference Reversal property is true, let us  suppose $a=\phi(\bar P)\neq\phi(\bar Q)=b$.

\medskip
Let us consider the function $v\mapsto (\bar P_v, \bar Q_v)$. Since its range is finite because of our assumption on $A$, let us list the elements of this range in the following way: $(W_1,W'_1),$ $(W_2, W'_2), \dots,$ $(W_\alpha, W'_\alpha)$ and the corresponding inverse images as
$V_1, V_2, ..., V_\alpha$.

For the society whose elements are $\{1, 2,\dots, \alpha\}$,  let us consider the scf 
$$\Phi:\{W_1, W'_1\}\times\{W_2,W'_2\}\times\dots\times\{W_\alpha, W'_\alpha\}\to A, \mbox{ in case (1) }$$
$$\Phi:\cO^\alpha\to A,  \mbox{ in case (2)}$$ 
 
 given as follows. From the profile $\pi=(\pi_{1},\dots,\pi_{\alpha})$, define the profile $P=(P_v)_{v\in V}$ by
$P_v=\pi_{i},\mbox{ for } v\in V_i$ and then, since $P\in\cP$ by construction,
$\Phi(\pi)=\phi (P).$ Notice that $\Phi$ is GSP, and  $$a=\phi(\bar P)=\Phi (W_1,W_2,\dots, W_\alpha)\mbox{ and } b=\phi(\bar Q)=\Phi (W'_1,W'_2,\dots, W'_\alpha).$$

\bigskip
Duly applying Theorem \ref{main} or Theorem \ref{New},  we get the Preference Reversal  property of $\Phi$. Hence there is an $i\le \alpha$ with
$W_i\neq W'_i$,\,\, $a\,\,\meglio_{W_i} b$, and  $ b\,\,\meglio_{W'_i} a$. 
Since for every $v\in V_i$ we have $(\bar P_v, \bar Q_v)
=(W_i, W'_i)$, we are done.
\eproof

\begin{oss}\label{final} {\rm As a final remark we note  that our analysis not only extends  \cite[Theorems 3 and 4]{BM} but also  provides some insights concerning the equivalence of individual and group strategy-proofness. Precisely:
\begin{itemize}

\item[-] The fact that in general  an ISP three-valued scf is GSP, is a consequence of the Preference Reversal property derived from the individual strategy-proofness. Compare with \cite[Corollary 1]{BBM},  where the authors go  through \cite[Proposition 1 and Theorem 1]{BBM} to prove the result.

\item[-] In the same way we show the equivalence under the Universal Domain hypothesis. Specifically: neither using the Gibbard-Satterthwaite Theorem\footnote{What in the weak order case, anyway requires an extra argument, as we have seen in Remark \ref{foot GS}.}   nor using the notion  of  a domain 
satisfying the {\it indirect sequential inclusion} property  (\cite [Section 4.4]{BBM} combined with  \cite[Theorem 2]{BBM}).
\item[-] We can go deeper  regarding this point.  Indeed, for proving the sufficiency of ISP for the Preference Reversal property, a weaker than Universal Domain hypothesis can be used. As proved by Theorem \ref{New}, it is enough to assume that the feasible set of orderings of every voter is complete, in the sense fixed by the Definition \ref{ED}. In this case again individual implies group strategy-proofness. 
\end{itemize}
}\end{oss}

\section{Conclusions}\label{conclusioni}

For a social choice function, we have analyzed the  relation between   non-manipulability and preference reversal.  Our analysis extends the recent findings of Berga and Moreno \cite{BM} and points out the role of individual strategy-proofness in deriving the preference reversal property. At the same time we have clarified that the strictness of the preferences does not play a role in the findings, whereas it is the finiteness or not of the set of agents that determines the use of individual versus group strategy-proofness. The fact that we were able to use individual rather than group strategy-proofness has also allowed to get immediately some of the known, and some new, equivalences between individual and group strategy-proofness. Further we have characterized group strategy-proof by introducing the notion of almost preference reversal property.

\end{document}